\pdfoutput=1

\documentclass[11pt]{article}
\usepackage{acl}
\usepackage{times}
\usepackage{latexsym}
\usepackage{amsmath}
\usepackage[T1]{fontenc}
\usepackage[utf8]{inputenc}
\usepackage{microtype}
\usepackage{inconsolata}
\usepackage{graphicx}

\definecolor{blizzardblue}{rgb}{0.67, 0.9, 0.93}
\usepackage{helvet}
\usepackage{courier} 
\usepackage{multirow}
\usepackage{xcolor}
\usepackage{booktabs}
\usepackage{adjustbox}
\usepackage{xcolor,colortbl}
\definecolor{bubbles}{rgb}{0.91, 1.0, 1.0}

\definecolor{Gray}{gray}{0.85}
\definecolor{LightCyan}{rgb}{0.88,1,1}
\definecolor{lightthulianpink}{rgb}{0.9, 0.56, 0.67}
\definecolor{mypink3}{cmyk}{0, 0.7808, 0.4429, 0.1412}
\newcolumntype{a}{>{\columncolor{Gray}}c}
\newcolumntype{b}{>{\columncolor{white}}c}

\newcommand\blfootnote[1]{%
\begingroup
\renewcommand\thefootnote{}\footnote{#1}%
\addtocounter{footnote}{-1}%
\endgroup
}

\title{LLM-Guided Multi-View Hypergraph Learning for Human-Centric Explainable Recommendation}

\author{
    Zhixuan Chu$^{1*}$ \text{,} \textbf{Yan Wang}$^{1*}$ \text{,} \textbf{Qing Cui}$^{1}$ \text{,} \textbf{Longfei Li}$^{1}$ \text{,} \\ \textbf{Wenqing Chen}$^{2\dagger}$ \text{,}  \textbf{Zhan Qin}$^{3}$, \textbf{Kui Ren}$^{3}$\\
    $^1$ \text{Ant Group}  $^2$ \text{Sun Yat-sen University} $^3$ \text{Zhejiang University}\\
    \text{\{chuzhixuan.czx, luli.wy, cuiqing.cq, longyao.llf\}@antgroup.com} \\
    \text{chenwq95@mail.sysu.edu.cn, \{qinzhan, kuiren\}@zju.edu.cn} \\
}

\begin{document}
\maketitle

\blfootnote{* Equal contribution.}
\blfootnote{$\dagger$ Corresponding author.}

\begin{abstract}
As personalized recommendation systems become vital in the age of information overload, traditional methods relying solely on historical user interactions often fail to fully capture the multifaceted nature of human interests. To enable more human-centric modeling of user preferences, this work proposes a novel explainable recommendation framework, i.e., LLMHG, synergizing the reasoning capabilities of large language models (LLMs) and the structural advantages of hypergraph neural networks. By effectively profiling and interpreting the nuances of individual user interests, our framework pioneers enhancements to recommendation systems with increased explainability. We validate that explicitly accounting for the intricacies of human preferences allows our human-centric and explainable LLMHG approach to consistently outperform conventional models across diverse real-world datasets. The proposed plug-and-play enhancement framework delivers immediate gains in recommendation performance while offering a pathway to apply advanced LLMs for better capturing the complexity of human interests across machine learning applications.
\end{abstract}

\section{Introduction}

Personalized recommendation systems have become indispensable tools, helping users discover content attuned to their unique preferences. However, accurately modeling the multifaceted nature of human interests remains an open challenge. Conventional methods like collaborative and content-based filtering struggle to fully capture the complex evolution of individual user inclinations. To enable the next generation of human-centric and interpretable recommenders, we need novel approaches that profile and reason about the intricacies of personal preferences to produce tailored and satisfactory suggestions. 

The advent of large language models (LLMs) \citep{brown2020language,openai2023gpt4,touvron2023llama,guan2023intelligent,chu2024professional,guan2023intelligent} presents an unparalleled opportunity to delve deeper into user behavior and preferences, promising to revolutionize recommendation systems through enhanced understanding and prediction capabilities \citep{wang2023enhancing,chu2023leveraging}. LLMs contain an abundance of world knowledge about items, concepts, and their interrelationships, acquired through ingesting vast swathes of data during pre-training. By utilizing the semantic reasoning capabilities of LLMs \citep{chu2023data}, we can effectively extract, tease apart, and comprehend the multitude of factors governing an individual's interests. For instance, LLMs can deduce nuanced preferences such as favored genres, themes, eras, and styles by analyzing an individual's past movie-watching records. The powerful relational reasoning capacities of LLMs further allow linking extracted preference facets to item characteristics and properties \cite{wan2024bridging}. As a result, LLMs facilitate constructing intricate user profiles encapsulating their multifarious and fluid interests — overcoming the limitations of previous works reliant solely on past user interactions \citep{wu2019session,wang2021session,li2023next,xue2023prompt,xue2023easytpp}. By complementing LLMs' understanding of users with analysis of historical sequences, more insightful and adaptive recommendation systems can be realized.

However, the integration of LLMs into recommendation frameworks poses a challenge, in that LLMs must handle the extraction and interpretation of user preferences from sparse and noisy data. Users may engage with a wide array of content, making it difficult to discern underlying patterns or interests. In this paper, we present a novel approach that synergizes the semantic depth of LLMs with the structural advantages of hypergraph neural networks to create a personalized recommendation system. By tapping into the aptitude of LLMs to comprehend semantic relationships within data, we obtain a rich set of angles that encapsulate a user's interests, i.e., Interest Angles (IAs), which are structured representations of user preferences, based on their behavioral history. With IAs as anchors, we utilize the abundant knowledge of LLM to categorize movies into multiple categories within each IA. These two steps can initially construct a hypergraph to roughly represent the user's preferences. To further refine the constructed hypergraph, we apply hypergraph structure learning techniques, allowing us to re-weight hyperedges. The resulting compact hypergraph representation, optimized to focus on the most salient aspects of the user's preferences, offers an accurate substrate for generating recommendations. Finally, we integrate this hypergraph embedding with latent embeddings obtained from the conventional sequential recommendation model. Our contributions are threefold \cite{wang2024llmrg}:

(1) We propose a plug-and-play recommendation enhancement framework, LLMHG, that utilizes LLMs to enable more human-centric recommendation systems.

(2) To the best of our knowledge, this work pioneers multi-view hypergraphs powered by LLMs to encapsulate the multitude of factors governing human interests. 

(3) We propose a strategy of hypergraph structure optimizations to refine LLM-based user profiling. By treating LLMs as expert but imperfect feature extractors and correcting their reasoning gaps, we enhance the quality of LLM outputs.

\section{Related Work}
\paragraph{Hypergraph and Recommendation.}
The relations between items under certain intents represent higher-order information. To accurately model these complex relationships, \textbf{\textit{hypergraphs}} can be incorporated into recommendation systems. Hypergraphs are a generalization of graphs in which a hyperedge is an arbitrary non-empty subset of the vertex set. Currently, there are three main types of hypergraphs. \textbf{\textit{Transition hyperedges}} \citep{hidasi2015session,wu2019session} directly encodes the sequential order of item transitions. \textbf{\textit{Context hyperedges}} \citep{wang2021session} capture local interests by applying a sliding window to item sequences. \textbf{\textit{Intent hyperedges}} \citep{li2023next} identify intent-specific associations by calculating similarities between prototype vectors for intents and item embeddings. These existing techniques rely on ``algorithmically'' extractive operations on behavioral data to construct hypergraphs. While surface-level statistics are uncovered, the higher-order semantics between items and intents are not truly obtained. Our work pioneers the use of LLMs to generate hypergraph structures. Rather than purely algorithmic computations, rich latent connections are synthesized through the expansive knowledge encoded within the parameters of the language model. This allows our method to produce a more holistic hypergraph containing nuanced semantic representations of the relationships.

\begin{figure*}[t]
    \centering
    \includegraphics[width=1.8\columnwidth]{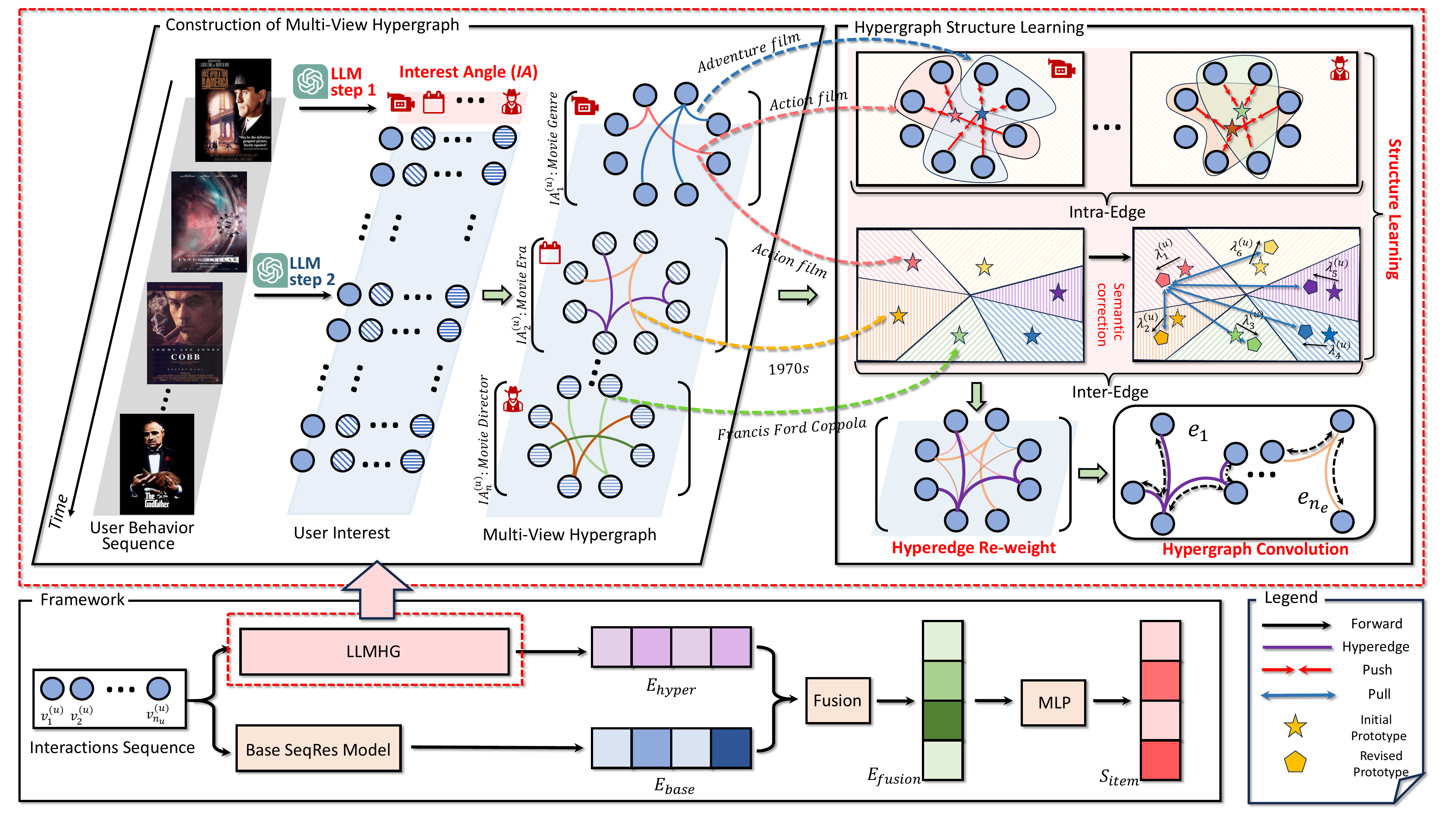}
    \caption{LLMHG includes four major steps: interest angle extraction, construction of a multi-view hypergraph centered on interest angles, hypergraph structure learning for LLM content refinement, and representation fusion for recommendation prediction.}
    \label{fig:framework}
\end{figure*}

\paragraph{LLMs and Recommendation.}
The integration of LLMs into recommendation systems is an emerging area of research that has shown considerable promise, according to pioneering studies. The potential of these models can be categorized into three distinct approaches, each harnessing the power of LLMs in innovative ways. \textbf{\textit{Using LLMs Directly for Recommendations.}} A novel aspect of LLMs is that they can be used in recommendation systems without constructing new models from scratch. These methods depend on crafting specific prompts for the LLMs \citep{liu2023chatgpt,gao2023chat,dai2023uncovering,chen2023palr} or applying minimal fine-tuning to adapt the model to the task of recommendations \citep{zhang2023recommendation,kang2023llms,bao2023tallrec}. \textbf{\textit{LLMs as Sources of Supplementary Information.}} LLMs act as sophisticated feature extractors that process information about items and users, subsequently producing context-rich embeddings \citep{wu2021empowering,qiu2021u,yao2022reprbert,muhamed2021ctr,xiao2022training,liu2023first,wang2022towards,wang2023generative}. These embeddings can then be seamlessly integrated into traditional recommendation models, thereby enriching them with the LLM's extensive knowledge base. \textbf{\textit{LLMs as Interactive Agents in Recommendation Systems.}} LLMs take a more active role by managing the entire recommendation process. These advanced models are adapted for use in recommendation contexts, where they can take charge of gathering user data, engineering features, encoding this information, and even directing the scoring and ranking mechanisms \citep{andreas2022language,bao2023tallrec,hou2023large,lin2023can,gao2023chat,friedman2023leveraging}. Together, these forward-looking approaches demonstrate the transformative potential of LLMs in revolutionizing recommendation systems.

\section{Methodology}

\subsection{Problem Statement}

Recommender systems aim to capture users' interests based on their historical interactions, viewing the user's history of interactions as an ordered sequence and attempting to model the user's dynamically evolving interests. Formally, let ${U}{=}\{u_1,u_2,\dots,u_{\vert{U}\vert}\}$ denote the set of users, ${V}{=}\{v_1,v_2,\dots,v_{|{V}|}\}$ be the set of items with the corresponding item attributes $\{x_1,x_2,\dots,x_{|{V}|}\}$, and list ${S}_u{=}[v_1^{(u)},\dots,v_t^{(u)},\dots,v_{n_u}^{(u)}]$ denote the sequence of interactions for user $u \in {U}$ in chronological order, where $v_t^{(u)} \in {V}$ is the item interacted with at time step $t$ and $n_u$ is the length of the sequence. We use relative time indices instead of absolute timestamps. Given a user's interaction history ${S}_u$, the sequential recommendation task is to predict the item the user $u$ will interact with at the next time step $n_{u} + 1$. This can be formalized as modeling the probability distribution over all possible items for user $u$ at time step $n_{u} {+} 1$, i.e., $p\bigl(v_{n_u+1}^{(u)}=v|\ {S}_u)$.

In this paper, we present a novel LLM-induced multi-view hypergraph recommendation approach, i.e., LLMHG, which utilizes the vast world knowledge of LLM and structure optimization of multi-view hypergraph to capture the user's interests implied in the historical behavior sequence. As shown in Figure \ref{fig:framework}, we detail our methodology in four major steps: interest angle extraction, construction of a multi-view hypergraph centered on interest angles, hypergraph structure learning for LLM content refinement, and representation fusion for recommendation prediction.

\subsection{Interest Angle Generation}
Instead of having the LLM directly unearth all complex relationships within the raw behavior data, we opt for a more guided two-step approach. The first critical phase relies on leveraging LLMs to deduce Interest Angles (IAs) - structured encapsulations of a user's preference facets extracted from their historical sequences. Here, we take the movie recommendation task as an example. First, the LLM analyzes the behavioral sequence to produce a list of Interest Angles, each representing a salient aspect that potentially governs the user’s preferences such as favored genres, directors, themes, eras, or countries of origin. As shown in Figure \ref{fig:framework}, using these extracted angles as anchors to guide detailed classification significantly boosts profiling accuracy compared to directly querying the LLM in one shot.

Furthermore, the discrete Interest Angles returned by the LLM lend themselves to straightforward integration with structured graphical constructs. These IAs constitute the basic building blocks for assembling a multi-view hypergraph, where each view houses hyperedges of movies that pertain to a specific angle. Together, they encapsulate the multitude of factors that influence an individual’s movie preferences from various perspectives. 

\subsection{Construction of Multi-View Hypergraph}
With the Interest Angles extracted, the second step is to leverage the extensive world knowledge encoded within the large language model to categorize movies from the user's behavioral history into specific groups along each angle. For instance, the model can map movies into finer genres like horror, action, drama, comedy, etc. if the ``genre'' angle has been extracted previously. This categorization facilitates assembling a multi-view hypergraph, with each view corresponding to one of the extracted angles (e.g. genre, director, country). The hyperedges within each view capture clusters of movies that share common attributes pertinent to that angle, based on the LLM's categorization. For example, the ``genre'' view may contain multiple genre-based hyperedges like ``sci-fi movies'', ``comedy'', ``romance movies'', and so on, which group together movies of those genres watched by the user. Essentially, each view and its hyperedges provide a projection of the user's movie-watching patterns along a specific facet. By collectively accounting for all extracted angles governing their interests, the multi-view hypergraph offers a comprehensive representation of a user's movie preferences. It profiles their inclination towards movies through multiple lenses - genres, themes, eras, etc. 

Specifically, we give some formal definitions for the constructed hypergraph. Given a hypergraph $G=(V,E)$,  the item set $V$ is the vertex set and the category set $E$ under all interest angles is the hyperedge set, where each vertex and hyperedge is respectively defined as $v\in V$ and $e \in E$. Moreover, a hyperedge $e$ is a subset of $V$, which means these items belong to the same category $e$ of a certain interest angle. The vertex-edge incidence matrix $H\in \mathcal{R}^{|V|\times |E|}$ is defined as follows:
\begin{equation*}
\small
h(v,e)=\left\{
\begin{array}{l l}
{1,~\text{if}~v\in e}\\
{0,~\text{otherwise}}.
\end{array}
\right.
\end{equation*}
The degree of a hyperedge $e$ is the number of vertices in $e$, i.e., $\delta(e)=\sum_{v\in e}h(v,e)$, and the degree of a vertex $v\in V$ is defined as $d(v)=\sum_{v\in e, e\in E} w(e)=\sum_{e\in E} w(e)h(v,e)$, where $w(e)$ is the weight of the hyperedge $e$. We denote the diagonal matrix forms of $\delta(e)$, $d(v)$ and $w(e)$ as $D_e$, $D_v$ and $W$ respectively. 

\subsection{Hypergraph Structure Learning for LLM Content Refinement}

The initial extraction of Interest Angles and the following classification may not fully capture the intricacies of human preferences. Two primary factors can limit the sufficiency: (1) Knowledge limitations. Despite their vast scope, LLMs may still lack information about all item concepts, attributes, and relationships. Their knowledge is fully dependent on the training data of LLM. As a result, they may fail to extract all critical preference facets. (2) Reasoning errors. LLMs remain fundamentally probabilistic systems, and their reasoning processes can occasionally propagate inaccuracies or logical gaps resulting in suboptimal Interest Angle extraction. To ameliorate the effects of such LLM limitations, instead of further fine-tuning which can be inefficient, we opt for post-processing through hypergraph structure learning including intra-edge and inter-edge structure learning. 

First, we compute a prototype $P(e)$, which is the centroid for each hyperedge (subcategory of interest angle). When a hyperedge contains a sufficient number of items, the average value among the items can serve as the prototype; however, when there are fewer items within a hyperedge, taking the average directly may lead to a deviation from the actual prototype and fail to fully express the prototype's information. Therefore, we utilize the text information (the specific categories of interest angle) generated by LLMs to supplement and correct the prototype of the hyperedge. The calculation process for the prototype is as follows:
\begin{equation*}
\begin{aligned}
P^{ori}(e_k) &= \frac{1}{| e_k |} \sum_{v_i \in e_k} f(v_i) \\
  p(e_k) &= (1 - \lambda_k) \cdot P^{ori}(e_k) +  \lambda_k \cdot T_{e_k} \\
  \lambda_k &= \frac{exp(-h(T_{e_k}))}{1 + exp(-h(T_{e_k}))},
\end{aligned}
\end{equation*}
where $P^{ori}(e_k)$ refers to the initial prototype for hyperedge $e_k$ computed directly using the average of the feature space. $p(e_k)$ is the prototype supplemented with the text information by LLM. $T_{e_k}$ is the embedding corresponding to the text information associated with $e_k$. The balance between $P^{ori}(e_k)$ and $T_{e_k}$ is adjusted through a parameter $\lambda_k$. $h(\cdot)$ is the learnable function in the coefficient $\lambda_k$ that integrates the text information.

Then, we regard the hyperedge $e$ as a clique and consider the mean of the heat kernel weights of the pairwise edges and the distance between prototypes as the hyperedge weight:
\begin{align*}
\textstyle w(e)= &\textstyle\frac{\beta}{\delta(e) (\delta(e) -1)} \sum\limits_{\{ v_i, v_j \} \in e} exp ( - \frac{\| \varphi(x_i) - \varphi(x_j) \|^2}{\mu})\\ 
& \textstyle +(1- \beta) \sum\limits_{e_k \in E} \frac{ \| P(e) - P(e_k) \|^2}{n_e} .
\end{align*}
This hyperedge weight function consists of two parts, where the first part is intra-edge structure learning, assessing the aggregation degree of items within a hyperedge; the second item is inter-edge structure learning which evaluates the sparsity of different hyperedge prototypes. We expect that the items within the same hyperedge can aggregate as much as possible, and there should be sparse between different hyperedges. Here, $\varphi(x_i)$ is a learnable kernel function and $n_e$ is the number of hyperedges, which is the sum of the numbers of subcategories in all interest angles.

A collection of hypergraph cuts $F=[f_1,\cdots,f_{n_e}]$ is defined as the predictors of hyperedges. Ideally, a cut in a hypergraph partition should minimize the disturbance to hyperedges. This means that the best cut aims to preserve the connections among items to the greatest extent, given that every hyperedge represents a category label. Similar to normalized hypergraph \cite{zhou2006learning}, the structure learning loss function given hypergraph $G$ and hyperedge cuts $F$ can be defined as \cite{huang2015learning}:

{\small\begin{align*}\nonumber
  \mathcal{L}_{str}(F, G)& =\frac{1}{2}\sum_{e\in E}\sum_{(v_i,v_j)\in e}\frac{w(e)}{\delta(e)}\left|\left|\frac{F_{v_i}}{\sqrt{d({v_i})}}-\frac{F_{v_j}}{\sqrt{d({v_j})}}\right|\right|^2\\
  &  =\text{Tr}(F^T(I-D_{v_j}^{-1/2}{HW D_e^{-1}H^T}{D_{v_j}^{-1/2}})F)\\
  &  =\text{Tr}(F^TL_HF),
\end{align*}}
\noindent where $F_{v_i}$ returns a vector representing the prediction of hyperedges for the item $v_i$. $L_H$ is the normalized hypergraph Laplacian matrix and $I$ is an identity matrix. $\text{Tr}(\cdot)$ is the trace of the matrix. 

For now, we can refine the structure of this hypergraph to suppress irrelevant facets and amplify salient preference associations. In essence, the hypergraph refinement stage acts as an inference correction mechanism approximating the ideal structure fully representative of an individual's tastes. Using this strategy of LLM extraction followed by structured post-inference adjustment, we can benefit both from the efficiency of LLMs for expansive knowledge extraction and the transparency of graphical learning methods - sidestepping shortcomings associated with both approaches.

\subsection{Representation Fusion}
In the final step, we integrate the representations derived from hypergraph structure learning with baseline representations obtained from a sequence recommendation model. Under the guidance of hyperedge weight $w(e)$, we adopt the hypergraph neural networks, which employ a hyperedge convolution operation to capture high-order relationships. The loss functions are defined as follows:

{\small\begin{align*}
 \mathcal{L}_{pre} &  = -\sum_{u=1}^{\vert\mathcal{U}\vert} y_u log( \hat{y}_u) + (1 - y_u) log(1 - \hat{y}_u) \\
 \mathcal{L} & = \mathcal{L}_{str} + \alpha \mathcal{L}_{pre},
\end{align*}}
where $y_u$ is the next item to be predicted for user $u$, and $\alpha$ is the hyperparameter to balance the hypergraph structure learning loss and the supervised prediction loss.

By fusing both representations, we aim to enhance the recommendation system's ability to predict the next item. This fusion creates a more explainable and accurate recommendation system that takes into account both the multifaceted nature of user preferences and the sequential behavior exhibited in behavioral history.

\begin{figure*}[t]
    \centering
    \includegraphics[width=1.9\columnwidth]{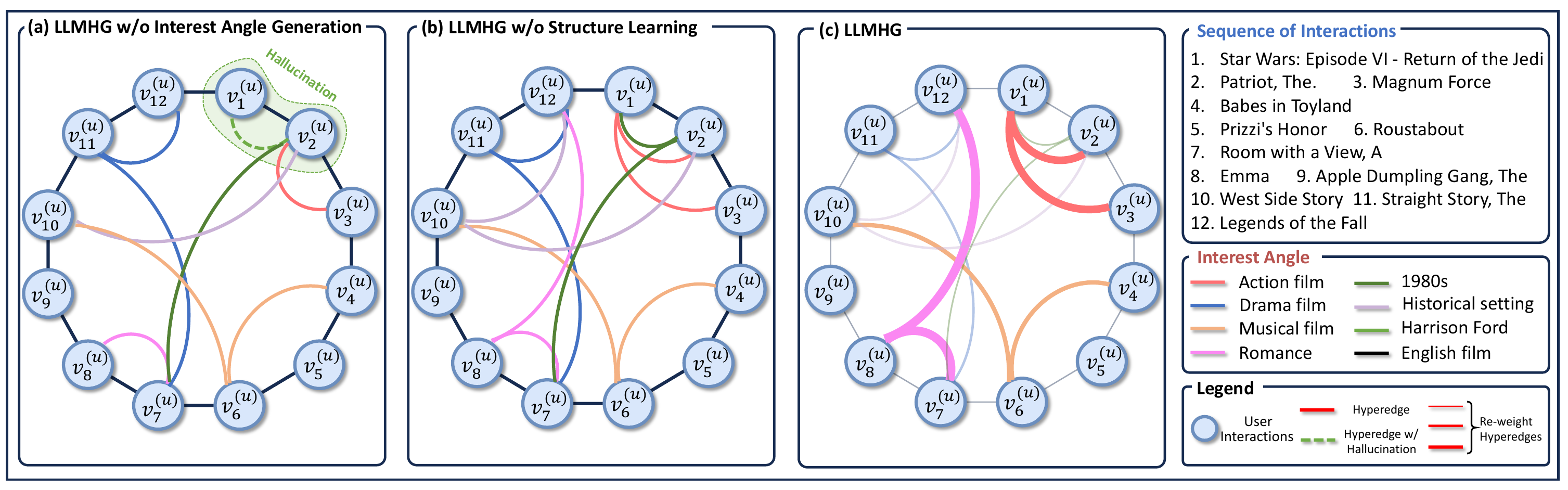}
    
    \caption{The real case studies (ML-1M) on our (c) LLMHG and ablation models, i.e., (a) LLMHG w/o interest angle generation  and (b) LLMHG w/o hypergraph structure learning .}
    
    \label{fig:real_case_example}
\end{figure*}

\section{Experiments}

\subsection{Experiment Settings}

\paragraph{Dataset.} To evaluate our proposed method, we conduct experiments on three benchmark datasets: the Amazon Beauty, Amazon Toys, and MovieLens-1M (ML-1M) datasets. The Amazon datasets, originally introduced in \citealp{mcauley2015image}, are known for high sparsity and short sequence lengths. We select the Beauty and Toys subcategories, using the fine-grained product categories and brands as item attributes. The ML-1M dataset, from \citep{harper2015movielens}, is a large and dense dataset consisting of long item sequences collected from the movie recommendation site MovieLens, with movie genres used as attributes. The statistics of the three datasets after preprocessing are summarized in Table \ref{Datasets}. Following common practice \citep{kang2018self,qiu2022contrastive,sun2019bert4rec,zhou2020s3}, we treat all user-item interactions as implicit feedback. For each user, we remove duplicate interactions and sort the remaining interactions chronologically to construct sequential user profiles. Through experiments on these diverse public datasets, we aim to thoroughly evaluate the performance of our proposed approach. 

\begin{table}

    \centering
    \caption{Statistics of the datasets after preprocessing.}
    
    \resizebox{0.8\columnwidth}{!}{
        \begin{tabular}{l| ccc}
        \toprule
        Specs. & Beauty & Toys  & ML-1M\\
        \midrule
       \# Users & 22,363 & 19,412  &  6,041 \\
       \# Items & 12,101 & 11,924  & 3,417 \\
       \# Avg.Length & 8.9 & 8.6  & 165.5\\
       \# Actions & 198,502 & 167,597  & 999,611\\
       Sparsity & 99.93\% & 99.93\%  & 95.16\%\\
       \bottomrule
    \end{tabular}}
    
    \label{Datasets}
\end{table}

\paragraph{Evaluation Metrics.}
To evaluate the performance of our recommendation system, we utilize a leave-one-out strategy where we repeatedly hold out one item from each user's sequence of interactions. This allows us to test the model's ability to predict the held-out item. We make predictions over the entire item set without any negative sampling. We report two widely used ranking metrics - Top-$n$ metrics \textbf{HR@$n$} (Hit Rate) and \textbf{NDCG@$n$} (Normalized Discounted Cumulative Gain) where $n$ is set to 5 and 10. HR@$n$ measures whether the held-out item is present in the top-$n$ recommendations, while NDCG@$n$ considers the position of the held-out item by assigning higher scores to hits at the top ranks. To ensure robust evaluation, we repeat each experiment 5 times with different random seeds and report the average performance across runs as the final metrics. This allows us to account for variability and ensure our results are not dependent on a particular random initialization.

\paragraph{Baselines.}
Following the experiment comparison in \citealp{du2023ensemble}, we include baseline methods from three groups for comparison: (1) General sequential methods utilize a sequence encoder to generate the hidden representations of users and items. For example, BERT4Rec \citep{sun2019bert4rec} adopts a bidirectional Transformer as the sequence encoder. (2) Attribute-aware sequential methods fuse attribute information into sequential recommendations. For example, FDSA \citep{zhang2019feature} applies self-attention blocks to capture transition patterns of items and attributes. (3) Contrastive sequential methods design auxiliary objectives for contrastive learning based on general sequential methods. For example, CL4SRec \citep{xie2022contrastive} proposes data augmentation strategies for contrastive learning in the sequential recommendation. DuoRec \citep{qiu2022contrastive} proposes both supervised and unsupervised sampling strategies for contrastive learning in the sequential recommendation.

\begin{table*}[h!]
    \centering
    
    \caption{Performance comparison on three benchmark datasets, i.e., ML-1M, Amazon Beauty, and Amazon Toys. We set the original models as baselines to compare with our proposed LLMHG model based on GPT3.5 or GPT4. The shaded area indicates the improved performance of our LLMHG model over the baselines across all three datasets. Higher is better.}
    
    \resizebox{1.9\columnwidth}{!}{
        \begin{tabular}{l l|baa |baa |baa |baa}
        \toprule
        \multirow{2.5}{*}{Dataset} & \multirow{2.5}{*}{Metric} &
        \multicolumn{3}{c}{FDSA} & \multicolumn{3}{c}{BERT4Rec} & 
        \multicolumn{3}{c}{CL4SRec} & \multicolumn{3}{c}{DuoRec} \\
        
          \cmidrule(lr){3-5} \cmidrule(lr){6-8} \cmidrule(lr){9-11} \cmidrule(lr){12-14}
        
          &  & Original  & GPT3.5 & GPT4 
         & Original  & GPT3.5 & GPT4
         & Original  & GPT3.5 & GPT4
         & Original  & GPT3.5 & GPT4\\
       
        \midrule
         \multirow{4}{4em}{ML-1M} & HR@5 & 0.0912  &  \textcolor{red}{+ 16.88\%} & \textcolor{red}{+ 20.72\%} & 0.1135 & \textcolor{red}{+ 14.62\%} & \textcolor{red}{+ 18.23\%} & 0.1147 & \textcolor{red}{+ 13.77\%} & \textcolor{red}{+ 18.04\%} & 0.2016 & \textcolor{red}{+ 8.97\%}&  \textcolor{red}{+ 10.81\%}\\
         & HR@10 &0.1644 & \textcolor{red}{+ 13.03 \%} & \textcolor{red}{+ 15.29\%} & 0.0.1917 & \textcolor{red}{+ 10.27 \%} & \textcolor{red}{+ 11.26 \%} & 0.1861 & \textcolor{red}{+ 11.92 \%} & \textcolor{red}{+ 17.57 \%} & 0.2840 & \textcolor{red}{+ 7.67 \%} & \textcolor{red}{+ 9.26 \%} \\
        & NDCG@5 & 0.0587 & \textcolor{red}{+ 14.65\%} & \textcolor{red}{+ 17.37 \%} & 0.0715 & \textcolor{red}{+ 9.37 \%} & \textcolor{red}{+ 12.16 \%} & 0.0714 & \textcolor{red}{+ 11.76 \%} & \textcolor{red}{+ 15.12 \%} & 0.1264 & \textcolor{red}{+ 7.91 \%} & \textcolor{red}{+ 10.12 \%} \\
        & NDCG@10 & 0.0874 & \textcolor{red}{+ 11.78\%} & \textcolor{red}{+ 14.64\%} & 0.0983 & \textcolor{red}{+ 8.54\%} & \textcolor{red}{+ 14.34\%} & 0.1010 & \textcolor{red}{+ 7.82\%} & \textcolor{red}{+ 10.29\%} & 0.1669 & \textcolor{red}{+ 6.35\%} & \textcolor{red}{+ 8.02\%} \\
         \midrule
        & HR@5 & 0.0230 & \textcolor{red}{+ 14.78 \%} & \textcolor{red}{+ 16.95 \%} & 0.0189 & \textcolor{red}{+ 13.22 \%} & \textcolor{red}{+ 16.40 \%} & 0.0391 & \textcolor{red}{+ 10.99 \%} & \textcolor{red}{+  12.53 \%} & 0.0547 & \textcolor{red}{+ 10.23 \%} & \textcolor{red}{+ 11.70 \%} \\
        Amazon & HR@10 & 0.0411  & \textcolor{red}{+ 12.41 \%} & \textcolor{red}{+ 15.08 \%} & 0.0401 & \textcolor{red}{+ 10.47 \%} & \textcolor{red}{+ 13.21 \%} & 0.0661 & \textcolor{red}{+ 8.87 \%} & \textcolor{red}{+ 10.89 \%} & 0.0835 & \textcolor{red}{+ 7.06 \%} & \textcolor{red}{+ 9.34 \%} \\
        Beauty & NDCG@5 & 0.0192 & \textcolor{red}{+ 9.37 \%} & \textcolor{red}{+ 13.02 \%} & 0.0188 & \textcolor{red}{+ 9.57 \%} & \textcolor{red}{+ 13.29 \%} & 0.0215 & \textcolor{red}{+ 6.97 \%} & \textcolor{red}{+ 9.76 \%} & 0.0344 & \textcolor{red}{+ 7.26 \%} & \textcolor{red}{+ 9.59 \%} \\
        & NDCG@10 & 0.0266 & \textcolor{red}{+ 8.64 \%} & \textcolor{red}{+ 10.52 \%} & 0.0260 & \textcolor{red}{+ 7.30 \%} & \textcolor{red}{+ 8.84 \%} & 0.0316 & \textcolor{red}{+ 6.32 \%} & \textcolor{red}{+ 9.81 \%} & 0.0431 & \textcolor{red}{+ 6.72 \%} & \textcolor{red}{+ 8.35 \%} \\
        \midrule
        & HR@5 &  0.0278 & \textcolor{red}{+ 13.66 \%} & \textcolor{red}{+ 17.62 \%} & 0.0379 & \textcolor{red}{+ 8.97 \%} & \textcolor{red}{+ 11.34 \%} & 0.0512 & \textcolor{red}{+ 10.74 \%} & \textcolor{red}{+ 14.84 \%} & 0.0531 & \textcolor{red}{+ 9.98 \%} & \textcolor{red}{+ 12.42 \%} \\
       Amazon  & HR@10 & 0.0501 & \textcolor{red}{+ 11.17 \%} & \textcolor{red}{+ 15.16 \%} & 0.0533 & \textcolor{red}{+ 11.06 \%} & \textcolor{red}{+ 12.94 \%} & 0.0729 & \textcolor{red}{+ 8.23 \%} & \textcolor{red}{+ 9.19 \%} & 0.0750 & \textcolor{red}{+ 6.80 \%}  &  \textcolor{red}{+ 8.53 \%}  \\
    Toys & NDCG@5 & 0.0216 & \textcolor{red}{+ 5.55 \%} & \textcolor{red}{+ 10.64 \%} & 0.0261 & \textcolor{red}{+ 9.96 \%} & \textcolor{red}{+ 11.11 \%} & 0.0266 & \textcolor{red}{+ 6.76 \%} & \textcolor{red}{+ 11.27 \%} & 0.0342 & \textcolor{red}{+ 6.14 \%} & \textcolor{red}{+ 10.81 \%} \\
        & NDCG@10 & 0.0283 & \textcolor{red}{+ 5.65 \%} & \textcolor{red}{+ 7.42 \%} & 0.0313 & \textcolor{red}{+ 7.02  \%} & \textcolor{red}{+ 9.26 \%} & 0.0341 & \textcolor{red}{+ 6.74 \%} & \textcolor{red}{+ 9.97 \%} & 0.0411 & \textcolor{red}{+ 6.32 \%} & \textcolor{red}{+ 8.27 \%} \\
        
       \bottomrule
    \end{tabular}}
    
    \label{Main_results}
\end{table*}

\begin{table*}[h!]
    \centering
    
    \caption{Ablation studies of our LLMHG model on two benchmark datasets, i.e., ML-1M and Amazon Beauty.}
    
    \resizebox{1.85\columnwidth}{!}{
        \begin{tabular}{l l |cccc |cccc }
        \toprule
       \multirow{2.5}{*}{LLM} & \multirow{2.5}{*}{Method} & \multicolumn{4}{c}{ML-1M} &
        \multicolumn{4}{c}{Amazon Beauty}  \\
        
          \cmidrule(lr){3-6} \cmidrule(lr){7-10} 
        
          & &  HR@5& HR@10  & NDCG@5 & NDCG@10 
         &  HR@5& HR@10  & NDCG@5 & NDCG@10\\
       
        \midrule
        NA & DuoRec & 0.2016 & 0.2840 & 0.1264 & 0.1669 & 0.0547 & 0.0835 & 0.0344 & 0.0431 \\
         \midrule
         \rowcolor{Gray}
         \multirow{6}{4em}{GPT3.5} & LLMHG(Ours) & \textbf{\textcolor{red}{+ 8.97 \%}} & \textbf{\textcolor{red}{+ 7.67 \%}} & \textbf{\textcolor{red}{+ 7.91 \%}} & \textbf{\textcolor{red}{+  6.35 \%}} & \textbf{\textcolor{red}{+ 10.23 \%}} & \textbf{\textcolor{red}{+ 7.06 \%}} & \textbf{\textcolor{red}{+ 7.26 \%}} & \textbf{\textcolor{red}{+ 6.72  \%}} \\
         & w/o IAs & \textcolor{red}{+ 5.33 \%} & \textcolor{red}{+ 4.87 \%} & \textcolor{red}{+ 4.79 \%} & \textcolor{red}{+ 3.96 \%} & \textcolor{red}{+ 6.56 \%} & \textcolor{red}{+ 4.97 \%} & \textcolor{red}{+ 5.33 \%} & \textcolor{red}{+ 4.94 \%} \\
        & w/o IntraSL & \textcolor{red}{+ 5.96 \%} & \textcolor{red}{+ 5.13 \%} & \textcolor{red}{+ 5.24 \%} & \textcolor{red}{+ 4.88 \%} & \textcolor{red}{+ 7.43 \%} & \textcolor{red}{+ 5.66 \%} & \textcolor{red}{+ 5.87 \%} & \textcolor{red}{+ 5.05 \%} \\
         & w/o InterSL & \textcolor{red}{+ 6.74 \%} & \textcolor{red}{+ 6.22 \%} & \textcolor{red}{+ 6.54 \%} & \textcolor{red}{+ 6.03 \%} & \textcolor{red}{+ 8.46 \%} & \textcolor{red}{+ 6.01 \%} & \textcolor{red}{+ 6.14 \%} & \textcolor{red}{+ 5.85 \%} \\
        & w/o ProCor  & \textcolor{red}{+ 8.54 \%} & \textcolor{red}{+ 7.13 \%} & \textcolor{red}{+ 7.66 \%} & \textcolor{red}{+ 6.21 \%} & \textcolor{red}{+ 9.76 \%} & \textcolor{red}{+ 6.75 \%} & \textcolor{red}{+ 6.72 \%} & \textcolor{red}{+ 6.31 \%} \\
        & w/o SL & \textcolor{red}{+ 3.15 \%} & \textcolor{red}{+ 2.99 \%} & \textcolor{red}{+ 3.26 \%} & \textcolor{red}{+ 2.91 \%} & \textcolor{red}{+ 5.54 \%} & \textcolor{red}{+ 4.78 \%} & \textcolor{red}{+ 4.96 \%} & \textcolor{red}{+ 4.34 \%} \\
       \midrule
       \rowcolor{Gray}
       \multirow{6}{4em}{GPT4} & LLMHG(Ours) & \textbf{\textcolor{red}{+ 10.81 \%}} & \textbf{\textcolor{red}{+ 9.26 \%}} & \textbf{\textcolor{red}{+ 10.12 \%}} & \textbf{\textcolor{red}{+ 8.02 \%}} & \textbf{\textcolor{red}{+ 11.70 \%}} & \textbf{\textcolor{red}{+ 9.34  \%}} & \textbf{\textcolor{red}{+ 9.59 \%}}& \textbf{\textcolor{red}{+ 8.35 \%}} \\
        & w/o IAs & \textcolor{red}{+ 8.96 \%} & \textcolor{red}{+ 8.01 \%} & \textcolor{red}{+ 8.56 \%} & \textcolor{red}{+ 7.65 \%} & \textcolor{red}{+ 9.14 \%} & \textcolor{red}{+ 8.76 \%} & \textcolor{red}{+ 8.89 \%} & \textcolor{red}{+ 8.03 \%} \\
        & w/o IntraSL & \textcolor{red}{+ 8.67 \%} & \textcolor{red}{+ 7.92 \%} & \textcolor{red}{+ 8.46  \%} & \textcolor{red}{+ 7.13 \%} & \textcolor{red}{+ 8.95 \%} & \textcolor{red}{+ 8.27 \%} & \textcolor{red}{+ 8.31 \%} & \textcolor{red}{+ 7.86 \%} \\
         & w/o InterSL & \textcolor{red}{+ 9.67 \%} & \textcolor{red}{+ 9.01 \%} & \textcolor{red}{+ 9.65 \%} & \textcolor{red}{+ 7.62 \%} & \textcolor{red}{+ 10.74 \%} & \textcolor{red}{+ 8.53 \%} & \textcolor{red}{+ 8.55 \%} & \textcolor{red}{+ 7.90 \%} \\
        & w/o ProCor  & \textcolor{red}{+ 10.66 \%} & \textcolor{red}{+ 9.14 \%} & \textcolor{red}{+ 10.06 \%} & \textcolor{red}{+ 7.93 \%} & \textcolor{red}{+ 11.22 \%} & \textcolor{red}{+ 8.99 \%} & \textcolor{red}{+ 9.10 \%} & \textcolor{red}{+ 8.07 \%} \\
        & w/o SL  & \textcolor{red}{+ 7.32 \%} & \textcolor{red}{+ 6.89 \%} & \textcolor{red}{+ 7.10 \%} & \textcolor{red}{+ 6.63 \%} & \textcolor{red}{+ 8.44 \%} & \textcolor{red}{+ 7.37 \%} & \textcolor{red}{+ 7.44 \%} & \textcolor{red}{+ 6.97 \%} \\

       \bottomrule
    \end{tabular}}
    
    \label{results_ablation}
\end{table*}

\subsection{Experiment Results}
As evidenced in Table \ref{Main_results}, we conduct comprehensive benchmarking experiments on three widely-used datasets - ML-1M, Amazon Beauty, and Amazon Toys. We compare our proposed LLMHG model built on top of GPT3.5 or GPT4 with several strong baseline methods, including FDSA \cite{zhang2019feature}, BERT4Rec \cite{sun2019bert4rec}, CL4SREC \cite{xie2022contrastive}, and DuoRec \cite{qiu2022contrastive}. The shaded regions in the table highlight the performance improvements achieved by our LLMHG model over all baselines across the three datasets. These results demonstrate the plug-and-play nature of LLMHG, which can effectively enhance multiple existing recommenders. This indicates that conventional recommender systems struggle to fully handle the extraction and interpretation of user preferences from sparse and noisy user behavioral history. These improvements showcase how large language models with hypergraph structure optimization can bring extensive background knowledge and strong logical reasoning to recommender systems. Furthermore, LLMHG performance scales with the underlying LLM capability - the GPT4-based LLMHG consistently outperforms its GPT3.5 counterpart.

\begin{figure*}[t]
    \centering
    \includegraphics[width=2\columnwidth]{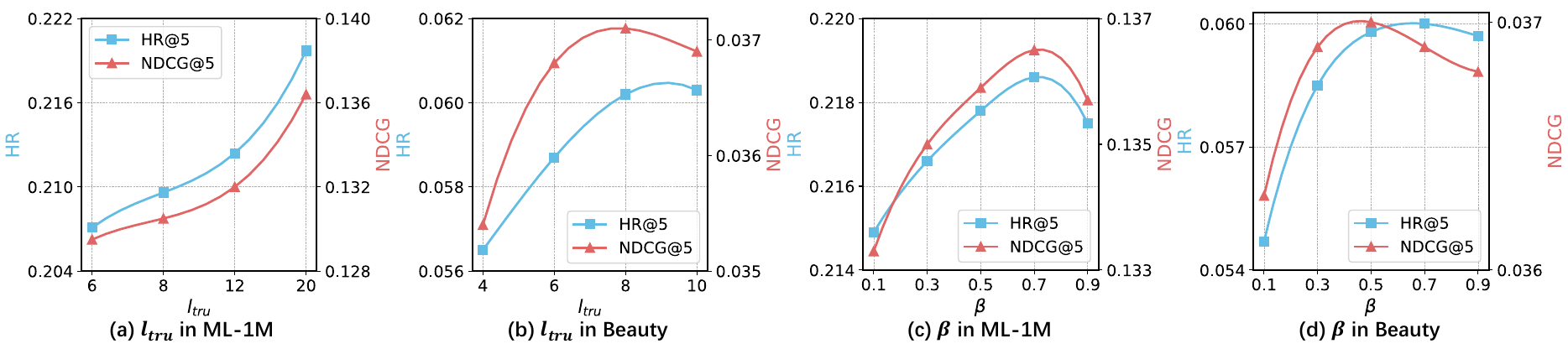}
    \vspace{-3mm}
    \caption{Sensitivity analysis of sequence truncation length $l_{tru}$ and intra and inter structure learning coefficient $\beta$ on HR and NDCG performance based on ML-1M and Amazon Beauty benchmarks.}
    \vspace{-3mm}
    \label{fig:exp_length_and_beta}
\end{figure*}

\begin{figure}[t]
    \centering
    \includegraphics[width=0.9\columnwidth]{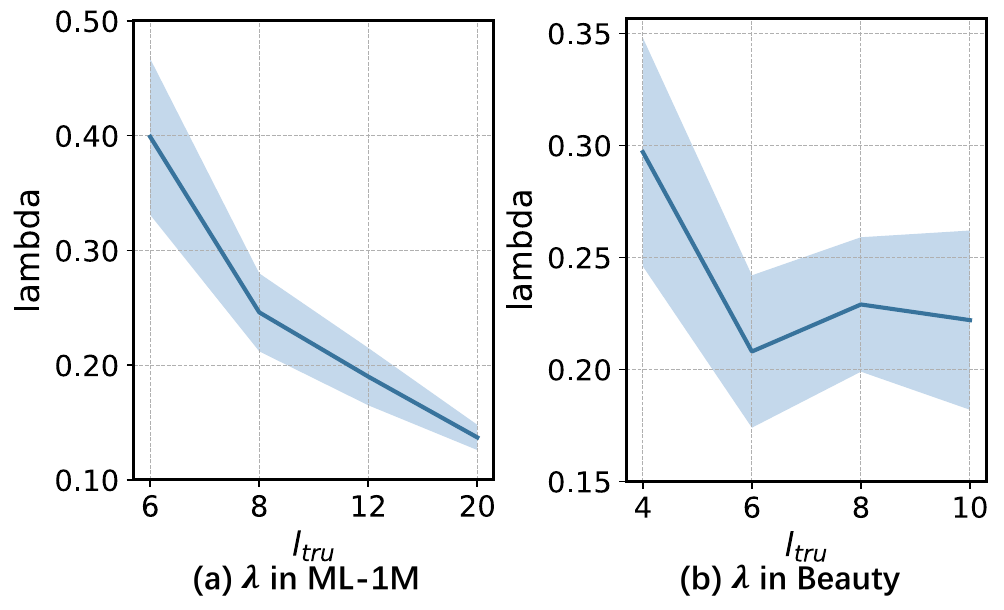}
    \vspace{-3mm}
    \caption{The analysis between prototype correction weight $\lambda$ and sequence truncation length $l_{tru}$.}
    \vspace{-7mm}
    \label{fig:exp_lambda}
\end{figure}

\subsection{Ablation Study}

To thoroughly evaluate each component of our proposed LLMHG framework, we conduct extensive ablation studies on the ML-1M and Amazon Beauty datasets. We systematically remove key modules of LLMHG and performance is compared to the best performance baseline model, i.e., DuoRec. Eliminating interest angle generation (LLMHG w/o IAs) directly constructs the hypergraph through LLM item categorization without the high-level guidance of extracted angles. Removing intra-edge structure learning (LLMHG w/o IntraSL) ablates the process of tightening clusters of related items within hyperedges. Disabling inter-edge structure learning (LLMHG w/o InterSL) eliminates the promotion of sparsity between diverse hyperedges capturing dissimilar preference facets. Excising prototype correction (LLMHG w/o ProCor) takes away the LLM-guided refinement of hyperedge centroids. Switching off all structure learning modules (LLMHG w/o SL) bypasses the complete graph optimization process and instead employs the hypergraph that is built directly from the original LLM output. As shown in Table \ref{results_ablation}, several insightful conclusions arise: (1) Each modular component positively contributes to overall performance - no element can be omitted without decreasing performance. This underscores the importance of the synergistic methodology: LLM knowledge extraction guided by interest angle and followed by hypergraph structure refinement. (2) Early interest angle deduction and later structure optimizations are more pivotal for the GPT3.5-based system compared to GPT4. This implies the inferior reasoning capabilities of GPT3.5 necessitate greater guidance and refinement. (3) The intra-edge process predominates over inter-edge learning. Homogenizing item groupings along a specific hyperedge seems more beneficial than simply separating between hyperedges. (4) Prototype correction effects are secondary, likely attributable to the sufficiency of item centroids derived purely from user history in representing hyperedges. (5) Notably, ablation harms are amplified on the ML-1M dataset compared to Amazon Beauty. The elaborate sequential patterns and multifaceted movie interests in ML-1M impose greater complexity - increasing dependency on interest angle guidance and structure optimizations to distill precise user profiles.

\begin{table*}[th!]
    \centering
    \caption{Comparison between LLMHG and DuoRec augmented with different hypergraphs, such as transition, context, and intent hyperedges, as well as DuoRec with the guidance of LLM.}
    \resizebox{1.9\columnwidth}{!}{
        \begin{tabular}{lll |cccc |cccc }
        \toprule
       & \multirow{2.5}{*}{Method} & & \multicolumn{4}{c}{ML-1M} &
        \multicolumn{4}{c}{Amazon Beauty}  \\
        
          \cmidrule(lr){4-7} \cmidrule(lr){8-11} 
        
         &&&  HR@5& HR@10  & NDCG@5 & NDCG@10 &
           HR@5& HR@10  & NDCG@5 & NDCG@10\\
       
        \midrule
         DuoRec  & & & 0.2016 & 0.2840 & 0.1264 & 0.1669 & 0.0547 & 0.0835 & 0.0344 & 0.0431 \\
         
         DuoRec & w/ & Transition HG & \textcolor{red}{+ 3.13 \%} & \textcolor{red}{+ 3.05 \%} & \textcolor{red}{+ 3.41 \%} & \textcolor{red}{+ 3.01 \%} & \textcolor{red}{+ 5.79 \%} & \textcolor{red}{+ 5.11 \%} & \textcolor{red}{+ 5.30 \%} & \textcolor{red}{+ 5.35 \%} \\
         DuoRec & w/ & Contextual HG & \textcolor{red}{+ 4.36 \%} & \textcolor{red}{+ 4.10 \%} & \textcolor{red}{+ 3.89 \%} & \textcolor{red}{+ 3.77 \%} & \textcolor{red}{+ 4.77 \%} & \textcolor{red}{+ 4.40 \%} & \textcolor{red}{+ 4.22 \%} & \textcolor{red}{+ 4.07 \%} \\
         DuoRec & w/ & Intent HG & \textcolor{darkgray}{- 2.52 \%} & \textcolor{darkgray}{- 2.37 \%} & \textcolor{darkgray}{- 2.79 \%} & \textcolor{darkgray}{- 2.30 \%} & \textcolor{red}{+ 1.03 \%} & \textcolor{red}{+ 0.91 \%} & \textcolor{red}{+ 0.87 \%} & \textcolor{red}{+ 0.69 \%} \\
         \rowcolor{blizzardblue}
         DuoRec+GPT3.5 &   &   & \textcolor{red}{+ 3.15 \% }& \textcolor{red}{+ 2.99 \% } & \textcolor{red}{+ 3.26 \% } & \textcolor{red}{+  2.91 \% } & \textcolor{red}{+ 5.54 \%}& \textcolor{red}{+ 4.78 \%} & \textcolor{red}{+ 4.96 \%} & \textcolor{red}{+ 4.34 \%} \\
        \rowcolor{Gray}
        LLMHG (GPT3.5) & &  & \textcolor{red}{+ 8.97 \%} & \textcolor{red}{+ 7.67 \%} & \textcolor{red}{+ 7.91 \%} & \textcolor{red}{+  6.35 \%} & \textcolor{red}{+ 10.23 \%} & \textcolor{red}{+ 7.06 \%} & \textcolor{red}{+ 7.26 \%} & \textcolor{red}{+ 6.72  \%} \\
       \rowcolor{blizzardblue}
        DuoRec+GPT4 &   &   & \textcolor{red}{+ 7.32  \%} & \textcolor{red}{+ 6.89 \%} & \textcolor{red}{+ 7.10 \%} & \textcolor{red}{+ 6.63 \%} & \textcolor{red}{+ 8.44 \%} & \textcolor{red}{+ 7.37 \%} & \textcolor{red}{+ 7.44 \%} & \textcolor{red}{+ 6.97 \%} \\
        \rowcolor{Gray}
        LLMHG (GPT4) & & & \textbf{\textcolor{red}{+ 10.81 \%}} & \textbf{\textcolor{red}{+ 9.26 \%}} & \textbf{\textcolor{red}{+ 10.12 \%}} & \textbf{\textcolor{red}{+ 8.02 \%}} & \textbf{\textcolor{red}{+ 11.70 \%}} & \textbf{\textcolor{red}{+ 9.34  \%}} & \textbf{\textcolor{red}{+ 9.59 \%}} & \textbf{\textcolor{red}{+ 8.35 \%}} \\

       \bottomrule
    \end{tabular}}
    \label{results_graph}
\end{table*}

\begin{table*}
    \centering
    
    \caption{Performance comparison experiments of LLMHG built on varying sized and different language models.}
    
    \resizebox{1.85\columnwidth}{!}{
        \begin{tabular}{l l |cccc |cccc }
        \toprule
       \multirow{2.5}{*}{Method} & \multirow{2.5}{*}{LLM} & \multicolumn{4}{c}{ML-1M} &
        \multicolumn{4}{c}{Amazon Beauty}  \\
        
          \cmidrule(lr){3-6} \cmidrule(lr){7-10} 
        
          & &  HR@5& HR@10  & NDCG@5 & NDCG@10 
         &  HR@5& HR@10  & NDCG@5 & NDCG@10\\
       
        \midrule
        DuoRec & NA & 0.2016 & 0.2840 & 0.1264 & 0.1669 & 0.0547 & 0.0835 & 0.0344 & 0.0431 \\
         \midrule
         \multirow{6}{4em}{LLMHG} & w/ Llama2-7B & \textcolor{darkgray}{- 3.12 \%} & \textcolor{darkgray}{- 2.52 \%} & \textcolor{darkgray}{- 2.76 \%} & \textcolor{darkgray}{- 2.60 \%} & \textcolor{red}{+ 2.33 \%} & \textcolor{red}{+ 2.15 \%} & \textcolor{red}{+ 1.93 \%} & \textcolor{red}{+ 1.77 \%} \\
        & w/ Llama2-13B & \textcolor{red}{+ 1.58 \%} & \textcolor{red}{+ 1.40 \%} & \textcolor{red}{+ 1.44 \%} & \textcolor{red}{+  1.37 \%} & \textcolor{red}{+ 4.79 \%} & \textcolor{red}{+ 4.50 \%} & \textcolor{red}{+ 4.39 \%} & \textcolor{red}{+ 4.32 \%} \\
         & w/ Qwen-7B & \textcolor{darkgray}{- 2.78 \%} & \textcolor{darkgray}{- 2.56 \%} & \textcolor{darkgray}{- 2.50 \%} & \textcolor{darkgray}{- 2.37 \%} & \textcolor{red}{+ 2.12 \%} & \textcolor{red}{+ 1.83 \%} & \textcolor{red}{+ 1.70 \%} & \textcolor{red}{+ 1.74 \%} \\
        & w/ Qwen-14B  & \textcolor{red}{+ 1.03 \%} & \textcolor{red}{+ 0.85 \%} & \textcolor{red}{+ 0.73 \%} & \textcolor{red}{+ 0.60 \%} & \textcolor{red}{+ 4.31 \%} & \textcolor{red}{+ 4.02 \%} & \textcolor{red}{+ 4.11 \%} & \textcolor{red}{+ 3.80 \%} \\
        \rowcolor{Gray}
        & w/ GPT3.5 & \textcolor{red}{+ 8.97 \%} & \textcolor{red}{+ 7.67 \%} & \textcolor{red}{+ 7.91 \%} & \textcolor{red}{+  6.35 \%} & \textcolor{red}{+ 10.23 \%} & \textcolor{red}{+ 7.06 \%} & \textcolor{red}{+ 7.26 \%} & \textcolor{red}{+ 6.72  \%} \\
        \rowcolor{Gray}
        & w/ GPT4 & \textbf{\textcolor{red}{+ 10.81 \%}} & \textbf{\textcolor{red}{+ 9.26 \%}} & \textbf{\textcolor{red}{+ 10.12 \%}} & \textbf{\textcolor{red}{+ 8.02 \%}} & \textbf{\textcolor{red}{+ 11.70 \%}} & \textbf{\textcolor{red}{+ 9.34  \%}} & \textbf{\textcolor{red}{+ 9.59 \%}} & \textbf{\textcolor{red}{+ 8.35 \%}} \\
       \bottomrule
    \end{tabular}}
    
    \label{results_div_ver}
\end{table*}

\subsection{Sensitivity Analysis}
We analyze the impact of three vital hyperparameters through sensitivity studies: sequence truncation length $l_{tru}$, intra and inter-structure learning coefficient $\beta$, and prototype correction weight $\lambda$. Figure \ref{fig:exp_length_and_beta} (a) and (b) examines the performance curve on ML-1M and Amazon Beauty when varying the truncated sequence lengths. We find 15-20 in ML-1M and 6-8 in Beauty are optimal, balancing history coverage with efficiency. Next, tuning the $\beta$ in Figure \ref{fig:exp_length_and_beta} (c) and (d) reveals that tighter intra-structure learning is better. Finally, Figure \ref{fig:exp_lambda} shows the automatically learned $\lambda$ that weights LLM guidance for constructing prototype vectors versus the relative contributions of item centroids. We observe that datasets with longer behavioral sequences, like ML-1M, tend to have lower $\lambda$ values. This indicates less reliance on LLM cues for prototype correction when users have rich historical data.

\subsection{Comparison on Different Hypergraphs}

We compare LLMHG to DuoRec augmented with different hypergraphs \cite{li2022enhancing} as well as DuoRec informed by the LLM. (1) \textbf{\textit{Sequential Transition Hyperedges}}. The relative chronological order of item transitions is a crucial aspect for recommenders. To preserve the inherent order of these item transitions, we link sequential items with hyperedges. This connection method effectively captures the temporal sequence of item interactions, vital for understanding user behavior. (2) \textbf{\textit{Contextual Hyperedges}}. The sequential context within a user's session offers insights into their latent interests. By deploying a sliding window technique over the sequence of items,  we create hyperedges that connect the items within each window. As we vary the size of these windows, we can extract insights into a user's interests at different scales, which are then integrated to form a comprehensive view of local user interests. (3) \textbf{\textit{Intent-based Hyperedges}}. The similarity between items can fluctuate significantly depending on the user's intent. For example, a Nikon camera and a Canon lens might be grouped together when the focus is on acquiring professional photography gear. However, these items might not be linked to the same brand. The intent-based hyperedges can reflect the correlation between items under specific intents. We start by calculating the cosine similarity between items and predefined intent prototypes. Then, for each distinct intent, a hyperedge is generated by connecting the top-n items ranked by their similarity score. (4) \textbf{\textit{Incorporating the LLM technique into DuoRec}}. The model is enhanced by supplementary information derived from the LLM. This additional background knowledge is first converted into vector representations of words \cite{mikolov2013efficient}, which are then integrated with the original item representation vectors. 

First, most hypergraph augmentations lead to modest gains over vanilla DuoRec, validating the benefit of graph structures in capturing multifaceted user interests. However, intent-based hyperedges bring little improvement, with minor gains on Amazon Beauty but decreases on the ML-1M dataset. As a result, traditional hypergraph construction methods struggle to effectively generate intent hyperedges. In addition, merely providing LLM guidance to DuoRec lags significantly behind LLMHG. This highlights LLMHG's unique methodology - not just LLM augmentation of the recommender but also structure refinement of the initial LLM outputs. 

\subsection{Comparison on Different LLMs}

\begin{table}
    \centering
    
    \caption{Cost-effectiveness analysis of our LLMHG model on ML-1M benchmarks. Higher is better.}
    
    \resizebox{1.0\columnwidth}{!}{
        \begin{tabular}{ll| c |cc |cc }
        \toprule
       \multirow{2.5}{*}{LLM} & \multirow{2.5}{*}{Method} & \multirow{2.5}{*}{Cost(USD)} & \multicolumn{2}{c}{HR@10} &
        \multicolumn{2}{c}{NDCG@10}  \\
        
          \cmidrule(lr){4-5} \cmidrule(lr){6-7} 
        
          & &  & Imp.$(\%)$ & CIR$(\%)$ & Imp.$(\%)$ & CIR$(\%)$ \\
       
        \midrule
         &LLMHG w/o IAs & \textcolor{darkgray}{0.0081} & \textcolor{darkgray}{ 4.87 } & \textcolor{darkgray}{ 601.23} & \textcolor{darkgray}{3.96} & \textcolor{darkgray}{488.89} \\

        \multirow{-2}{4em}{GPT3.5}  & LLMHG (ours) & \textcolor{darkgray}{0.0141} & \textcolor{darkgray}{7.67} & \textcolor{darkgray}{543.97} & \textcolor{darkgray}{6.35} & \textcolor{darkgray}{450.35} \\
        \midrule
          & LLMHG w/o IAs & \textcolor{darkgray}{0.2681} & \textcolor{darkgray}{8.01} & \textcolor{darkgray}{29.87} & \textcolor{darkgray}{7.65} & \textcolor{darkgray}{28.53} \\

        \multirow{-2}{4em}{GPT4} & LLMHG (ours) & \textcolor{darkgray}{0.4936} & \textcolor{darkgray}{9.26} & \textcolor{darkgray}{18.76} & \textcolor{darkgray}{8.02 } & \textcolor{darkgray}{16.24} \\
       \bottomrule
    \end{tabular}}
    
    \label{results_cir}
\end{table}

We additionally conduct experiments comparing LLMHG when built on varying sized and different language models. Concretely, we evaluate Llama2-7B, Llama2-13B, Qwen-7B, and Qwen-14B as well as GPT3.5 and GPT4. Intriguingly, we find that when using smaller Llama2-7B/Qwen-7B models as the backbone, LLMHG underperforms the DuoRec baseline, decreasing on ML-1M and increasing slightly on Amazon Beauty. In contrast, based on larger Llama2-13B/Qwen-14B LLMs, LLMHG begins to offer more gains. This divergence likely arises as smaller models contain more factual inconsistencies and reasoning errors. In comparison, the GPT3.5 and GPT4 foundation enables LLMHG to achieve its best performance. In total, our additional LLM analysis highlights scaling model capacity as vital for fact-based reasoning in LLMHG. 

\subsection{The Analysis of LLM Consumption}

Model consumption is a critical consideration when deploying LLMs for personalized recommendation systems. We evaluate the cost-effectiveness of our LLMHG model across three key metrics: Cost, average improvement rate (Imp), and Cost and Improvement Rate (CIR). The ``Cost'' represents the expenditure of utilizing the LLMHG model for each user, accounting for the fees incurred by accessing the LLMs (GPT-3.5 Turbo and GPT-4) through the HuggingFace API. The ``Imp.'' denotes the average performance boost in HR@10 and NDCG@10 between the LLMHG and baseline DuoRec models. Finally, ``CIR'' measures the rate of performance improvement versus cost for each LLM configuration. The results in Table \ref{results_cir} show that while advancements such as GPT-4 and tailored interest generation result in significant performance boosts, they have an inverse effect on the cost and CIR, sharply increasing costs and reducing CIR. The impact is more pronounced with GPT-4. Therefore, more advanced LLMs can greatly improve recommendations but require more computational resources. This cost-accuracy analysis allows system designers to pick the optimal LLM that fits their use case, resources, and improvements sought.

\section{Limitation and Discussion}

Our pioneering integration of hypergraph techniques to refine LLM outputs enables a robust framework for improving generative quality. By treating large language models as intelligent but imperfect feature extractors and correcting their reasoning gaps via structured hypergraph inference, we provide a plug-and-play solution to enhance personalized recommendations and other tasks.

However, limitations remain in our current approach. The detached post-processing pipeline likely results in less tight integration versus a joint end-to-end training regimen between the LLM and hypergraph components. Architecting such a unified procedure could better optimize the full stack, further improving extracted interest facets and recommendations. Additionally, our generic hypergraph optimization algorithm does not specifically target the types of reasoning lapses - such as causal inconsistencies, missing relational inferences, contextual unawareness, or graph structural biases \cite{salewski2024context,chu2024task,long2023can,chu2021graph}- that surface in large language models. By tailoring the hypergraph update rules to these observed LLM failure modes, we could potentially achieve even greater gains.

\section{Conclusion}
In summary, our proposed framework facilitates nuanced LLM-based user profiling while still accounting for sequential user behavior. Through these innovative steps, our methodology advances the field of personalized recommendations by integrating expansive world knowledge and expert-level reasoning proficiency of large language models with the representational efficacy of hypergraphs. Through comprehensive experiments on real-world datasets, we have demonstrated the efficacy of our proposed methodology - significantly improving state-of-the-art baselines.

\bibliography{llmhg}

\appendix

\section{Appendix}

\begin{table*}[h!]
    \centering
    
    \caption{Performance comparison on three benchmark datasets, i.e., ML-1M, Amazon Beauty, and Amazon Toys. We set the original models as baselines to compare with our proposed LLMHG model based on GPT3.5 or GPT4. The shaded area indicates the improved performance of our LLMHG model over the baselines across all three datasets. Higher is better.}
    
    \resizebox{1.9\columnwidth}{!}{
        \begin{tabular}{l l|baa |baa |baa |baa}
        \toprule
        \multirow{2.5}{*}{Dataset} & \multirow{2.5}{*}{Metric} &
        \multicolumn{3}{c}{FDSA} & \multicolumn{3}{c}{BERT4Rec} & 
        \multicolumn{3}{c}{CL4SRec} & \multicolumn{3}{c}{DuoRec} \\
        
          \cmidrule(lr){3-5} \cmidrule(lr){6-8} \cmidrule(lr){9-11} \cmidrule(lr){12-14}
        
          &  & Original  & GPT3.5 & GPT4 
         & Original  & GPT3.5 & GPT4
         & Original  & GPT3.5 & GPT4
         & Original  & GPT3.5 & GPT4\\
       
        \midrule
         \multirow{4}{4em}{ML-1M} & HR@5 & 0.0912  &  \textcolor{darkgray}{0.1066} & \textcolor{darkgray}{0.1101} & 0.1135 & \textcolor{darkgray}{0.1301} & \textcolor{darkgray}{0.1342} & 0.1147 & \textcolor{darkgray}{0.1305} & \textcolor{darkgray}{0.1354} & 0.2016 & \textcolor{darkgray}{0.2197}&  \textcolor{darkgray}{0.2234}\\
         & HR@10 &0.1644 & \textcolor{darkgray}{0.1858} & \textcolor{darkgray}{0.1995} & 0.1917 & \textcolor{darkgray}{0.2114} & \textcolor{darkgray}{0.2133} & 0.1861 & \textcolor{darkgray}{0.2083} & \textcolor{darkgray}{0.2188} & 0.2840 & \textcolor{darkgray}{0.3058} & \textcolor{darkgray}{0.3103} \\
        & NDCG@5 & 0.0587 & \textcolor{darkgray}{0.0673} & \textcolor{darkgray}{0.0689} & 0.0715 & \textcolor{darkgray}{0.0782} & \textcolor{darkgray}{0.0802} & 0.0714 & \textcolor{darkgray}{0.0798} & \textcolor{darkgray}{0.0822} & 0.1264 & \textcolor{darkgray}{0.1364} & \textcolor{darkgray}{0.1392} \\
        & NDCG@10 & 0.0874 & \textcolor{darkgray}{0.0977} & \textcolor{darkgray}{0.1002} & 0.0983 & \textcolor{darkgray}{0.1067} & \textcolor{darkgray}{0.1124} & 0.1010 & \textcolor{darkgray}{0.1089} & \textcolor{darkgray}{0.1114} & 0.1669 & \textcolor{darkgray}{0.1775} & \textcolor{darkgray}{0.1803} \\
         \midrule
        & HR@5 & 0.0230 & \textcolor{darkgray}{0.0264} & \textcolor{darkgray}{0.0269} & 0.0189 & \textcolor{darkgray}{0.0214} & \textcolor{darkgray}{0.0220} & 0.0391 & \textcolor{darkgray}{0.0434} & \textcolor{darkgray}{0.0440} & 0.0547 & \textcolor{darkgray}{0.0603} & \textcolor{darkgray}{0.0611} \\
        Amazon & HR@10 & 0.0411  & \textcolor{darkgray}{0.0462} & \textcolor{darkgray}{0.0473} & 0.0401 & \textcolor{darkgray}{0.0443} & \textcolor{darkgray}{0.0454} & 0.0661 & \textcolor{darkgray}{0.0719} & \textcolor{darkgray}{0.0733} & 0.0835 & \textcolor{darkgray}{0.0894} & \textcolor{darkgray}{0.0913} \\
        Beauty & NDCG@5 & 0.0192 & \textcolor{darkgray}{0.0210} & \textcolor{darkgray}{0.0217} & 0.0188 & \textcolor{darkgray}{0.0206} & \textcolor{darkgray}{0.0213} & 0.0215 & \textcolor{darkgray}{0.0230} & \textcolor{darkgray}{0.0236} & 0.0344 & \textcolor{darkgray}{0.0369} & \textcolor{darkgray}{0.0377} \\
        & NDCG@10 & 0.0266 & \textcolor{darkgray}{0.0289} & \textcolor{darkgray}{0.0294} & 0.0260 & \textcolor{darkgray}{0.0279} & \textcolor{darkgray}{0.0283} & 0.0316 & \textcolor{darkgray}{0.0336} & \textcolor{darkgray}{0.0347} & 0.0431 & \textcolor{darkgray}{0.0460} & \textcolor{darkgray}{0.0467} \\
        \midrule
        & HR@5 &  0.0278 & \textcolor{darkgray}{0.0316} & \textcolor{darkgray}{0.0327} & 0.0379 & \textcolor{darkgray}{0.0413} & \textcolor{darkgray}{0.0422} & 0.0512 & \textcolor{darkgray}{0.0567} & \textcolor{darkgray}{0.0588} & 0.0531 & \textcolor{darkgray}{0.0584} & \textcolor{darkgray}{0.0597} \\
       Amazon  & HR@10 & 0.0501 & \textcolor{darkgray}{0.0557} & \textcolor{darkgray}{0.0577} & 0.0533 & \textcolor{darkgray}{0.0592} & \textcolor{darkgray}{0.0602} & 0.0729 & \textcolor{darkgray}{0.0789} & \textcolor{darkgray}{0.0796} & 0.0750 & \textcolor{darkgray}{0.0801}  &  \textcolor{darkgray}{0.0814}  \\
    Toys & NDCG@5 & 0.0216 & \textcolor{darkgray}{0.0228} & \textcolor{darkgray}{0.0239} & 0.0261 & \textcolor{darkgray}{0.0287} & \textcolor{darkgray}{0.0290} & 0.0266 & \textcolor{darkgray}{0.0284} & \textcolor{darkgray}{0.0296} & 0.0342 & \textcolor{darkgray}{0.0363} & \textcolor{darkgray}{0.0379} \\
        & NDCG@10 & 0.0283 & \textcolor{darkgray}{0.0293} & \textcolor{darkgray}{0.0304} & 0.0313 & \textcolor{darkgray}{0.0335} & \textcolor{darkgray}{0.0342} & 0.0341 & \textcolor{darkgray}{0.0364} & \textcolor{darkgray}{0.0375} & 0.0411 & \textcolor{darkgray}{0.0437} & \textcolor{darkgray}{0.0445} \\
        
       \bottomrule
    \end{tabular}}
    
    \label{Main_results_appendix}
\end{table*}

\begin{table*}[h!]
    \centering
    
    \caption{Ablation studies of our LLMHG model on two benchmark datasets, i.e., ML-1M and Amazon Beauty.}
    
    \resizebox{1.85\columnwidth}{!}{
        \begin{tabular}{l l |cccc |cccc }
        \toprule
       \multirow{2.5}{*}{LLM} & \multirow{2.5}{*}{Method} & \multicolumn{4}{c}{ML-1M} &
        \multicolumn{4}{c}{Amazon Beauty}  \\
        
          \cmidrule(lr){3-6} \cmidrule(lr){7-10} 
        
          & &  HR@5& HR@10  & NDCG@5 & NDCG@10 
         &  HR@5& HR@10  & NDCG@5 & NDCG@10\\
       
        \midrule
        NA & DuoRec & 0.2016 & 0.2840 & 0.1264 & 0.1669 & 0.0547 & 0.0835 & 0.0344 & 0.0431 \\
         \midrule
         \rowcolor{Gray}
         \multirow{6}{4em}{GPT3.5} & LLMHG(Ours) & \textbf{\textcolor{darkgray}{0.2197}} & \textbf{\textcolor{darkgray}{0.3058}} & \textbf{\textcolor{darkgray}{0.1364}} & \textbf{\textcolor{darkgray}{0.1775}} & \textbf{\textcolor{darkgray}{0.0603}} & \textbf{\textcolor{darkgray}{0.0894}} & \textbf{\textcolor{darkgray}{0.0369}} & \textbf{\textcolor{darkgray}{0.0460}} \\
         & w/o IAs & \textcolor{darkgray}{0.2123} & \textcolor{darkgray}{0.2978} & \textcolor{darkgray}{0.1325} & \textcolor{darkgray}{0.1735} & \textcolor{darkgray}{0.0583} & \textcolor{darkgray}{0.0876} & \textcolor{darkgray}{0.0362} & \textcolor{darkgray}{0.0452} \\
        & w/o IntraSL & \textcolor{darkgray}{0.2136} & \textcolor{darkgray}{0.2986} & \textcolor{darkgray}{0.1330} & \textcolor{darkgray}{0.1750} & \textcolor{darkgray}{0.0588} & \textcolor{darkgray}{0.0882} & \textcolor{darkgray}{0.0364} & \textcolor{darkgray}{0.0453} \\
         & w/o InterSL & \textcolor{darkgray}{0.2151} & \textcolor{darkgray}{0.3017} & \textcolor{darkgray}{0.1347} & \textcolor{darkgray}{0.1770} & \textcolor{darkgray}{0.0593} & \textcolor{darkgray}{0.0885} & \textcolor{darkgray}{0.0365} & \textcolor{darkgray}{0.0456} \\
        & w/o ProCor  & \textcolor{darkgray}{0.2188} & \textcolor{darkgray}{0.3042} & \textcolor{darkgray}{0.1361} & \textcolor{darkgray}{0.1773} & \textcolor{darkgray}{0.0600} & \textcolor{darkgray}{0.0891} & \textcolor{darkgray}{0.0367} & \textcolor{darkgray}{0.0458} \\
        & w/o SL & \textcolor{darkgray}{0.2080} & \textcolor{darkgray}{0.2924} & \textcolor{darkgray}{0.1305} & \textcolor{darkgray}{0.1718} & \textcolor{darkgray}{0.0577} & \textcolor{darkgray}{0.0875} & \textcolor{darkgray}{0.0361} & \textcolor{darkgray}{0.0450} \\
       \midrule
       \rowcolor{Gray}
       \multirow{6}{4em}{GPT4} & LLMHG(Ours) & \textbf{\textcolor{darkgray}{0.2234}} & \textbf{\textcolor{darkgray}{0.3103}} & \textbf{\textcolor{darkgray}{0.1392}} & \textbf{\textcolor{darkgray}{0.1803}} & \textbf{\textcolor{darkgray}{0.0611}} & \textbf{\textcolor{darkgray}{0.0913}} & \textbf{\textcolor{darkgray}{0.0377}}& \textbf{\textcolor{darkgray}{0.0467}} \\
        & w/o IAs & \textcolor{darkgray}{0.2197} & \textcolor{darkgray}{0.3067} & \textcolor{darkgray}{0.1372} & \textcolor{darkgray}{0.1797} & \textcolor{darkgray}{0.0597} & \textcolor{darkgray}{0.0908} & \textcolor{darkgray}{0.0375} & \textcolor{darkgray}{0.0466} \\
        & w/o IntraSL & \textcolor{darkgray}{0.2191} & \textcolor{darkgray}{0.3065} & \textcolor{darkgray}{0.1371} & \textcolor{darkgray}{0.1788} & \textcolor{darkgray}{0.0596} & \textcolor{darkgray}{0.0904} & \textcolor{darkgray}{0.0372} & \textcolor{darkgray}{0.0465} \\
         & w/o InterSL & \textcolor{darkgray}{0.2211} & \textcolor{darkgray}{0.3096} & \textcolor{darkgray}{0.1386} & \textcolor{darkgray}{0.1796} & \textcolor{darkgray}{0.0606} & \textcolor{darkgray}{0.0906} & \textcolor{darkgray}{0.0373} & \textcolor{darkgray}{0.0465} \\
        & w/o ProCor  & \textcolor{darkgray}{0.2231} & \textcolor{darkgray}{0.3040} & \textcolor{darkgray}{0.1391} & \textcolor{darkgray}{0.1801} & \textcolor{darkgray}{0.0608} & \textcolor{darkgray}{0.0910} & \textcolor{darkgray}{0.0375} & \textcolor{darkgray}{0.0466} \\
        & w/o SL  & \textcolor{darkgray}{0.2164} & \textcolor{darkgray}{0.3037} & \textcolor{darkgray}{0.1354} & \textcolor{darkgray}{0.1780} & \textcolor{darkgray}{0.0593} & \textcolor{darkgray}{0.0897} & \textcolor{darkgray}{0.0370} & \textcolor{darkgray}{0.0461} \\

       \bottomrule
    \end{tabular}}
    
    \label{results_ablation_appendix}
\end{table*}

\begin{table*}[th!]
    \centering
    \caption{Comparison between LLMHG and DuoRec augmented with different hypergraphs, such as transition, context, and intent hyperedges, as well as DuoRec with the guidance of LLM.}
    \resizebox{1.9\columnwidth}{!}{
        \begin{tabular}{lll |cccc |cccc }
        \toprule
       & \multirow{2.5}{*}{Method} & & \multicolumn{4}{c}{ML-1M} &
        \multicolumn{4}{c}{Amazon Beauty}  \\
        
          \cmidrule(lr){4-7} \cmidrule(lr){8-11} 
        
         &&&  HR@5& HR@10  & NDCG@5 & NDCG@10 &
           HR@5& HR@10  & NDCG@5 & NDCG@10\\
       
        \midrule
         DuoRec  & & & 0.2016 & 0.2840 & 0.1264 & 0.1669 & 0.0547 & 0.0835 & 0.0344 & 0.0431 \\
         
         DuoRec & w/ & Transition HG & \textcolor{darkgray}{0.2079} & \textcolor{darkgray}{0.2927} & \textcolor{darkgray}{0.1307} & \textcolor{darkgray}{0.1719} & \textcolor{darkgray}{0.0597} & \textcolor{darkgray}{0.0878} & \textcolor{darkgray}{0.0362} & \textcolor{darkgray}{0.0454} \\
         DuoRec & w/ & Contextual HG & \textcolor{darkgray}{0.2103} & \textcolor{darkgray}{0.2956} & \textcolor{darkgray}{0.1313} & \textcolor{darkgray}{0.1732} & \textcolor{darkgray}{0.0573} & \textcolor{darkgray}{0.0872} & \textcolor{darkgray}{0.0359} & \textcolor{darkgray}{0.0449} \\
         DuoRec & w/ & Intent HG & \textcolor{darkgray}{0.1965} & \textcolor{darkgray}{0.2773} & \textcolor{darkgray}{0.1229} & \textcolor{darkgray}{0.1631} & \textcolor{darkgray}{0.0553} & \textcolor{darkgray}{0.0843} & \textcolor{darkgray}{0.0347} & \textcolor{darkgray}{0.0434} \\
         \rowcolor{blizzardblue}
         DuoRec+GPT3.5 &   &   & \textcolor{darkgray}{0.2080}& \textcolor{darkgray}{0.2924} & \textcolor{darkgray}{0.1305 } & \textcolor{darkgray}{0.1718 } & \textcolor{darkgray}{0.0577}& \textcolor{darkgray}{0.0875} & \textcolor{darkgray}{0.0361} & \textcolor{darkgray}{0.0450} \\
        \rowcolor{Gray}
        LLMHG (GPT3.5) & &  & \textbf{\textcolor{darkgray}{0.2197}} & \textbf{\textcolor{darkgray}{0.3058}} & \textbf{\textcolor{darkgray}{0.1364}} & \textbf{\textcolor{darkgray}{0.1775}} & \textbf{\textcolor{darkgray}{0.0603}} & \textbf{\textcolor{darkgray}{0.0894}} & \textbf{\textcolor{darkgray}{0.0369}} & \textbf{\textcolor{darkgray}{0.0460}} \\
       \rowcolor{blizzardblue}
        DuoRec+GPT4 &   &   & \textcolor{darkgray}{0.2164} & \textcolor{darkgray}{0.3037} & \textcolor{darkgray}{0.1354} & \textcolor{darkgray}{0.1780} & \textcolor{darkgray}{0.0593} & \textcolor{darkgray}{0.0897} & \textcolor{darkgray}{0.0370} & \textcolor{darkgray}{0.0461} \\
        \rowcolor{Gray}
        LLMHG (GPT4) & & & \textbf{\textcolor{darkgray}{0.2234}} & \textbf{\textcolor{darkgray}{0.3103}} & \textbf{\textcolor{darkgray}{0.1392}} & \textbf{\textcolor{darkgray}{0.1803}} & \textbf{\textcolor{darkgray}{0.0611}} & \textbf{\textcolor{darkgray}{0.0913}} & \textbf{\textcolor{darkgray}{0.0377}}& \textbf{\textcolor{darkgray}{0.0467}} \\

       \bottomrule
    \end{tabular}}
    \label{results_graph_appendix}
\end{table*}

\begin{table*}
    \centering
    
    \caption{Performance comparison experiments of LLMHG built on varying sized and different language models.}
    
    \resizebox{1.85\columnwidth}{!}{
        \begin{tabular}{l l |cccc |cccc }
        \toprule
       \multirow{2.5}{*}{Method} & \multirow{2.5}{*}{LLM} & \multicolumn{4}{c}{ML-1M} &
        \multicolumn{4}{c}{Amazon Beauty}  \\
        
          \cmidrule(lr){3-6} \cmidrule(lr){7-10} 
        
          & &  HR@5& HR@10  & NDCG@5 & NDCG@10 
         &  HR@5& HR@10  & NDCG@5 & NDCG@10\\
       
        \midrule
        DuoRec & NA & 0.2016 & 0.2840 & 0.1264 & 0.1669 & 0.0547 & 0.0835 & 0.0344 & 0.0431 \\
         \midrule
         \multirow{6}{4em}{LLMHG} & w/ Llama2-7B & \textcolor{darkgray}{0.1953} & \textcolor{darkgray}{0.2768} & \textcolor{darkgray}{0.1229} & \textcolor{darkgray}{0.1626} & \textcolor{darkgray}{0.0560} & \textcolor{darkgray}{0.0853} & \textcolor{darkgray}{0.0351} & \textcolor{darkgray}{0.0439} \\
        & w/ Llama2-13B & \textcolor{darkgray}{0.2047} & \textcolor{darkgray}{0.2880} & \textcolor{darkgray}{0.1282} & \textcolor{darkgray}{0.1692} & \textcolor{darkgray}{0.0573} & \textcolor{darkgray}{0.0873} & \textcolor{darkgray}{0.0359} & \textcolor{darkgray}{0.0450} \\
         & w/ Qwen-7B & \textcolor{darkgray}{0.1960} & \textcolor{darkgray}{0.2767} & \textcolor{darkgray}{0.1232} & \textcolor{darkgray}{0.1629} & \textcolor{darkgray}{0.0559} & \textcolor{darkgray}{0.0850} & \textcolor{darkgray}{0.0350} & \textcolor{darkgray}{0.0438} \\
        & w/ Qwen-14B  & \textcolor{darkgray}{0.2937} & \textcolor{darkgray}{0.2864} & \textcolor{darkgray}{0.1273} & \textcolor{darkgray}{0.1679} & \textcolor{darkgray}{0.0571} & \textcolor{darkgray}{0.0869} & \textcolor{darkgray}{0.0358} & \textcolor{darkgray}{0.0448} \\
        \rowcolor{Gray}
        & w/ GPT3.5 & \textbf{\textcolor{darkgray}{0.2197}} & \textbf{\textcolor{darkgray}{0.3058}} & \textbf{\textcolor{darkgray}{0.1364}} & \textbf{\textcolor{darkgray}{0.1775}} & \textbf{\textcolor{darkgray}{0.0603}} & \textbf{\textcolor{darkgray}{0.0894}} & \textbf{\textcolor{darkgray}{0.0369}} & \textbf{\textcolor{darkgray}{0.0460}} \\
        \rowcolor{Gray}
        & w/ GPT4 & \textbf{\textcolor{darkgray}{0.2234}} & \textbf{\textcolor{darkgray}{0.3103}} & \textbf{\textcolor{darkgray}{0.1392}} & \textbf{\textcolor{darkgray}{0.1803}} & \textbf{\textcolor{darkgray}{0.0611}} & \textbf{\textcolor{darkgray}{0.0913}} & \textbf{\textcolor{darkgray}{0.0377}}& \textbf{\textcolor{darkgray}{0.0467}} \\
       \bottomrule
    \end{tabular}}
    
    \label{results_div_ver_appendix}
\end{table*}

\begin{figure*}[t]
    \centering
    \includegraphics[width=2\columnwidth]{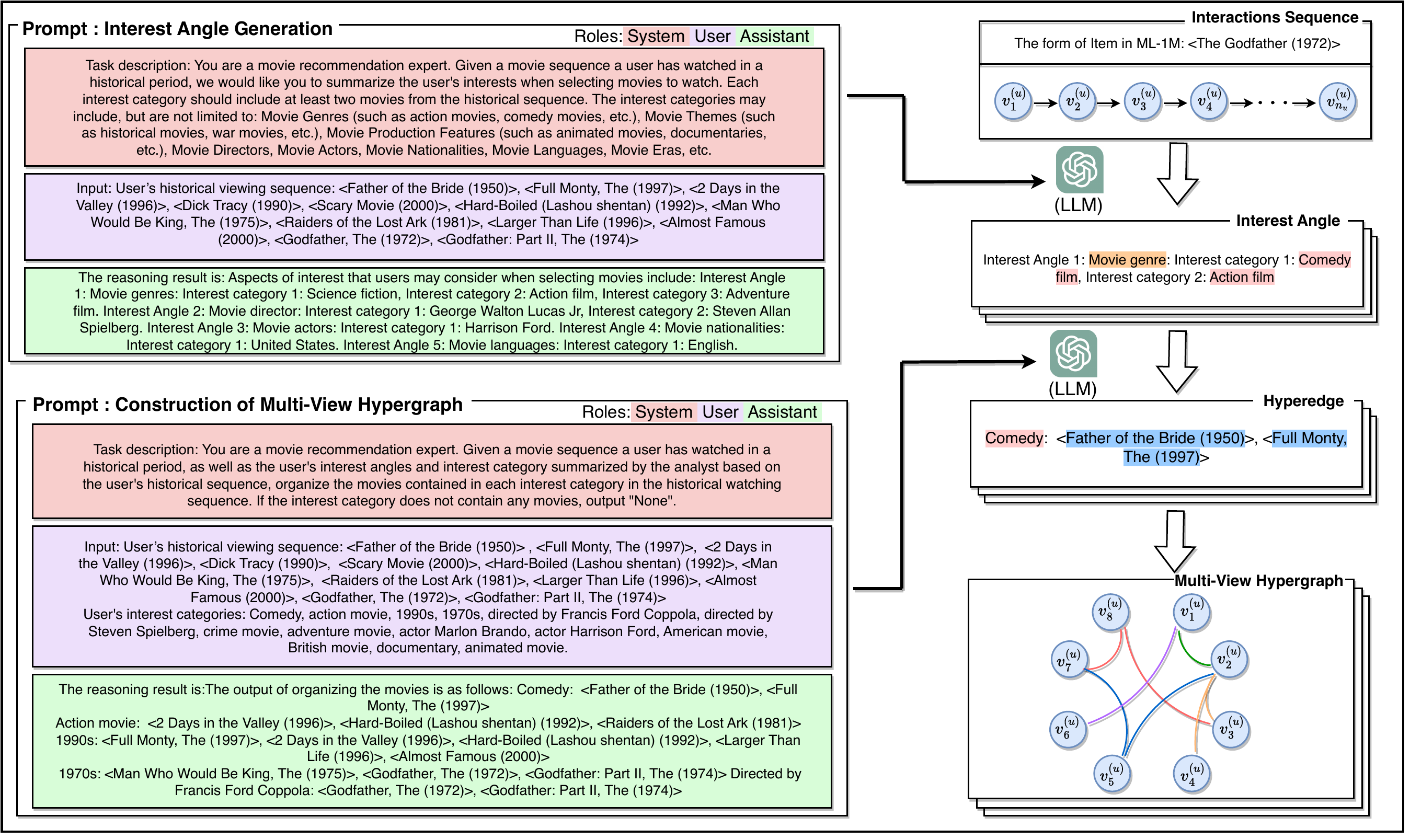}
    \caption{The prompt examples and real cases for LLMHG.}
    \label{fig:prompt_example}
\end{figure*}

\end{document}